\documentclass{article}
\usepackage{amssymb}
\usepackage{amsfonts}
\usepackage{amsmath}

\input{tcilatex}
\begin{document}

\title{On the Supergravitational Solitons}
\author{Vladimir A. Belinski \\
ICRANet, 65122 Pescara, Italy and\\
IHES, F-91440 Bures-sur-Yvette, France}
\maketitle

\begin{abstract}
The extention of the integrable ansatz of pure Einstein gravity to
supergravity is completed. The procedure of construction the exact
supergravitational solitonic solutions is described.
\end{abstract}

\section{Introduction}

The existence of the integrable ansatz in General Relativity has been
discovered long time ago \cite{M, BZ1}. In 1987 appeared the first version
of the integrable two-dimensional supergravity \cite{N1}. In the last paper
the maximal $N=16$ integrable supergravity in two space-time dimensions has
been considered and corresponding generalization of the Lax pair has been
proposed. The fact that in \cite{N1} only the maximal supergravity was
investigated is of no crucial significance because mathematics developed can
be adjusted to any $N.$ And, indeed, all principal points of the integrable
supergravity models was demonstrated in paper \cite{N2} using the simplest
case $N=2$.

However, these constructions have some shortcomings. First of all the Lax
pairs presented in \cite{N1, N2} are not complete since they contain (as
their self-consistency conditions) only bosonic part of the equations of
motion. The equations of motion for spinor fields do not follow from these
linear spectral problems and must be added by hands. Such mixed approach to
the integration can not be satisfactory in full. A manner how to apply it
for the construction of the exact solutions of the whole system of equations
of motion is intricate and such a way does not represents integrability in
the conventional sense. Another undesirable point which also creates some
non-standard complications is appearance in the linear spectral equations
the poles of the second order with respect to the spectral parameter while
the corresponding spectral problem in pure gravity has only simple poles.
One more circumstance we would like to mention is characteristic for many
papers dedicated to the integrable systems and not only to the articles \cite%
{N1, N2}. The point is that the authors often became fully satisfied as soon
as they showed the existence of the Lax pair and they do not pay attention
to the next even more important task: how to solve these equations. However,
to construct a procedure for extraction the exact solutions of the spectral
problem represents the main\textit{\ }part of the integration process.

In the present paper we will remove these drawbacks by extension the Lax
representation proposed in \cite{N1, N2} to the complete one (covering also
fermionic equation of motion) and deliberate it from the second order poles.
Also we will show how one can get the exact super-solitonic solutions using
this extended approach. We will restrict ourselves by the simplest case $N=2$
analyzed in paper \cite{N2} which, as we already said, does not represents
any principal loss of generality. The generalization of the approach
developed here for any $N$ is straightforward.

\section{ Integrable supergravity $N=2$ in two space-time dimensions}

The bosonic sector of the integrable theory presented in \cite{N2} consists
of the two-dimensional space-time\footnote{%
In our paper the three Greek indices $\lambda ,\mu ,\nu $ take only two
values $0$ and $1$. We introduce also the two-dimensional Minkowski metric
tensor $\eta _{\mu \nu }=diag(1,-1)$ then in the conformal gauge $j_{\mu \nu
}=$ $\eta _{\mu \nu }f$ $\ $the scale factor $f$ \ is positive.
\par
The simple partial derivatives are designated by comma. Tilde at the top of
a matrix means transposition. Matrices we designate by the bold letters. The
functions which depend on the spectral parameter we designate by the letters
with the hat on the top.} with interval $j_{\mu \nu }(x^{0},x^{1})dx^{\mu
}dx^{\nu }$ and of four real "matter" scalar fields living in this
two-dimensional space-time and depending only on coordinates $x^{0},x^{1}.$
These scalars is viewed as $2\times 2$ matrix $\mathbf{V}(x^{0},x^{1})$ and
the theory is \textit{postulated} to be invariant under the internal
symmetry transformation%
\begin{equation}
\mathbf{V}=GL(2)\cdot \mathbf{V}^{^{\prime }}\mathbf{\cdot }\text{ }O(2)%
\text{ },  \label{1}
\end{equation}%
where $GL(2)$ is rigid and $O(2)$\textbf{\ }is local (i.e. depending on $%
x^{0},x^{1}$). \ 

This system is the same as in paper \cite{BZ1}. Indeed, any $\mathbf{V}$
generates the symmetric matrix $\mathbf{G}$\textbf{:} 
\begin{equation}
\mathbf{G}=\mathbf{V\tilde{V}}\text{ },  \label{2}
\end{equation}%
which can be considered as metric tensor of some complementary
two-dimensional space-like manifold parametrized by dummy coordinates $%
x^{M}=(x^{2},x^{3})$ which\footnote{%
We use the uppercase Latin indices $\ $to correspond to these dummy
coordinates $x^{M}$ and these indices take only two values $2$ and $3.$ For
the corresponding frame indices in complementary space we use the same
letters but with the lines on the top. The mute coordinates $x^{2},x^{3}$
can be consider either pure imaginary (in fact this is our case) or real,
the choice depends on the signature of the whole four-dimensional \ metrics.
\par
{}} have nothing to do with $x^{\mu }=(x^{0},x^{1})$. The components of
matrices\footnote{%
In our notations for the matrix components the first (in the horizontal
direction) index, independent of its up or down position, enumerates the
rows and the second index (in the horizontal direction), independent of its
up or down position, enumerates the columns.} $\mathbf{G}$ and $\mathbf{V}$
are $G_{MN}$ and $V_{M\bar{N}}$ then the metric tensor of the complementary
space is $G_{MN}=\delta ^{\bar{J}\bar{L}}V_{M\bar{J}}V_{N\bar{L}}$ where $%
\delta ^{\bar{J}\bar{L}}$ signifies the Kronecker delta. Now the
four-dimensional interval 
\begin{equation}
ds^{2}=j_{\mu \nu }(x^{0},x^{1})dx^{\mu }dx^{\nu }+G_{MN}\left(
x^{0},x^{1}\right) dx^{M}dx^{N}  \label{3}
\end{equation}%
of paper \cite{BZ1} gives the geometrical representation of the system
considered in \cite{N2}. The Einstein-Hilbert Lagrangian (expressed in the
terms of $j_{\mu \nu }$ and $V_{M\bar{N}}$) for this interval is the same as
bosonic part of the Lagrangian of the paper \cite{N2}. The abstract
symmetries (\ref{1}) are realized in the interval (\ref{3}) as arbitrary
linear transformations with constant coefficients of the mute coordinates $%
x^{M}$ (this is $GL(2)$ acting on $\mathbf{V}$ from the left) and as local
(i.e. depending on the basic coordinates $x^{\mu }$) orthogonal rotations $%
O(2)$ of the frame acting on $\mathbf{V}$ from the right.

From the point of view of the two-dimensional space-time with coordinates $%
x^{0},x^{1}$ the metric coefficients $j_{\mu \nu }\left( x^{0},x^{1}\right) $
represent gravity and fields $V_{M\bar{N}}\left( x^{0},x^{1}\right) $
describe the bosonic "matter" filling this two-dimensional space. It is
known \cite{BZ1} that in the conformal gauge $j_{\mu \nu }=\eta _{\mu \nu }f$
$\ $equations of motion for the "matter" fields $V_{M\bar{N}}\left(
x^{0},x^{1}\right) $ are closed in the sense that they do not depend on the
conformal factor $f\left( x^{0},x^{1}\right) .$ This factor (which has no
propagating degrees of freedom of its own) follows from completely separate
system of the differential equations of the first order where from it can be
explicitly expressed in terms of the known solution for $V_{M\bar{N}}\left(
x^{0},x^{1}\right) .$ The essential result of the papers \cite{N1, N2} is
demonstration that also in supergravity exists the analogues super-gauge in
which two-dimensional interval has the form $j_{\mu \nu }=\eta _{\mu \nu }f$
and the aforementioned separable structure of the equations of motion
remains the same: in this super-gauge equations of motion for the scalar
fields $V_{M\bar{N}}\left( x^{0},x^{1}\right) $ and their superpartners form
the closed autonomous system of differential equations independent on the
conformal factor $f\left( x^{0},x^{1}\right) $ and of its fermionic
superpartners. The latter fields (all of which have no propagating degrees
of freedom of their own) can be explicitly expressed in terms of the former.
Then, exactly as in pure gravity, the only problem is to integrate the
equations of motion for the "matter" fields $V_{M\bar{N}}\left(
x^{0},x^{1}\right) $ together with their fermionic counterparts.

For the determinant of the frame matrix $\mathbf{V}$ we introduce the
notation: \ \ \ \ \ \ 
\begin{equation}
\det \mathbf{V=}\text{ }\alpha \text{ }.\text{ }  \label{4}
\end{equation}%
It turns out that, as in pure gravity, $\alpha $ satisfy the free wave
equation in the two-dimensional space-time $\left( x^{0},x^{1}\right) $ and
the super-gauge proposed in \cite{N1, N2} has the preference that \textit{%
the superpartners of the field }$\alpha $\textit{\ is eliminated} by the
allowable local supersymmetry transformation. This greatly relieves the
procedure of integration. Using the still permissible conformal
transformation of coordinates $x^{\mu }$ the function $\alpha $ can be
chosen just as one of the coordinates $x^{0}$ or $x^{1}$. We will keep $%
\alpha $ in its general unspecified form remembering, however, that it is a
non-physical field. Then the scalars $V_{M\bar{N}}\left( x^{0},x^{1}\right) $
contain only two bosonic "matter" physical degrees of freedom (these are
components of the symmetric matrix $\mathbf{G}$ subordinated to the
restriction $\det \mathbf{G}$ $=\alpha ^{2}$). For $N=2$ supergravity the
fermionic "matter" sector consists of two two-component (\textit{%
anticommuting}) Majorana fermions $\chi ^{\left( 1\right) }\left(
x^{0},x^{1}\right) $ and $\chi ^{\left( 2\right) }\left( x^{0},x^{1}\right) $
but on-shell they contain only two physical degrees of freedom (off-shell
the system need two auxiliary bosonic fields). The equations of motion for
this system have been written in \cite{N2} with the help of the
decomposition of the matrix current $\mathbf{V}^{-1}\mathbf{V}_{,\mu }$ into
its antisymmetric and symmetric parts: 
\begin{equation}
\mathbf{V}^{-1}\mathbf{V}_{,\mu }=\mathbf{Q}_{\mu }+\mathbf{P}_{\mu }\text{ }%
,\text{ }\mathbf{\tilde{Q}}_{\mu }=-\mathbf{Q}_{\mu }\text{ },\text{ }%
\mathbf{\tilde{P}}_{\mu }=\mathbf{P}_{\mu }\text{ }.  \label{5}
\end{equation}%
In two dimensions the matrix basis can be chosen as%
\begin{equation}
\mathbf{I=}\left( 
\begin{array}{cc}
1 & 0 \\ 
0 & 1%
\end{array}%
\right) \text{ },\text{ }\mathbf{Y}^{1}=\left( 
\begin{array}{cc}
1 & 0 \\ 
0 & -1%
\end{array}%
\right) \text{ },\text{ }\mathbf{Y}^{2}=\left( 
\begin{array}{cc}
0 & 1 \\ 
1 & 0%
\end{array}%
\right) \text{ },\text{ }\mathbf{Y}^{3}=\left( 
\begin{array}{cc}
0 & 1 \\ 
-1 & 0%
\end{array}%
\right) \text{ }.  \label{6}
\end{equation}%
Then matrices $\mathbf{P}_{\mu }$ and $\mathbf{Q}_{\mu }$ are:%
\begin{equation}
\mathbf{P}_{\mu }=p_{\mu }^{\left( 1\right) }\mathbf{Y}^{1}+p_{\mu }^{\left(
2\right) }\mathbf{Y}^{2}+\alpha _{,\mu }\left( 2\alpha \right) ^{-1}\mathbf{I%
}\text{ },\text{ }\mathbf{Q}_{\mu }=q_{\mu }\mathbf{Y}^{3}  \label{7}
\end{equation}%
with coefficients $p,q$ depending on coordinates $x^{0},x^{1}$ [the last
term in the matrix $\mathbf{P}_{\mu }$ is due to the trace of equation (\ref%
{5}) and notation (\ref{4})]. In terms of them the equations of motion
presented in \cite{N2} take the following form\footnote{%
In paper \cite{N2} there are misprints: both right hand sides of equations (%
\ref{9}) have there opposite signs and both right hand sides of equations (%
\ref{10}) contain the additional factor ($-2/3$).} :%
\begin{equation}
\eta ^{\mu \nu }\alpha _{,\mu \nu }=0\text{ },  \label{8}
\end{equation}%
\begin{eqnarray}
\alpha ^{-1}\eta ^{\mu \nu }\left[ \left( \alpha p_{\mu }^{\left( 1\right)
}\right) _{,\nu }+2q_{\nu }\alpha p_{\mu }^{\left( 2\right) }\right]
&=&-3ip_{\mu }^{\left( 2\right) }\bar{\chi}^{\left( 1\right) }\gamma ^{\mu
}\chi ^{\left( 2\right) }\text{ },  \label{9} \\
\alpha ^{-1}\eta ^{\mu \nu }\left[ \left( \alpha p_{\mu }^{\left( 2\right)
}\right) _{,\nu }-2q_{\nu }\alpha p_{\mu }^{\left( 1\right) }\right]
&=&3ip_{\mu }^{\left( 1\right) }\bar{\chi}^{\left( 1\right) }\gamma ^{\mu
}\chi ^{\left( 2\right) }\text{ },  \notag
\end{eqnarray}%
and%
\begin{eqnarray}
2i\gamma ^{\mu }\left[ \left( \sqrt{\alpha }\chi ^{\left( 1\right) }\right)
_{,\mu }+2q_{\mu }\sqrt{\alpha }\chi ^{\left( 2\right) }\right] &=&3\left( 
\bar{\chi}^{\left( 2\right) }\chi ^{\left( 2\right) }\right) \sqrt{\alpha }%
\chi ^{\left( 1\right) }\text{ },  \label{10} \\
2i\gamma ^{\mu }\left[ \left( \sqrt{\alpha }\chi ^{\left( 2\right) }\right)
_{,\mu }-2q_{\mu }\sqrt{\alpha }\chi ^{\left( 1\right) }\right] &=&3\left( 
\bar{\chi}^{\left( 1\right) }\chi ^{\left( 1\right) }\right) \sqrt{\alpha }%
\chi ^{\left( 2\right) }\text{ }.  \notag
\end{eqnarray}%
Here $\gamma ^{\mu }=\eta ^{\mu \nu }\gamma _{\nu }$ (that is $\gamma
^{0}=\gamma _{0},$ $\gamma ^{1}=-\gamma _{1}$) and matrices $\gamma _{\nu
}^{{}}$ are flat. They are chosen as%
\begin{equation}
\gamma _{0}=\left( 
\begin{array}{cc}
0 & -i \\ 
i & 0%
\end{array}%
\right) \text{ },\text{ }\gamma _{1}=\left( 
\begin{array}{cc}
-i & 0 \\ 
0 & i%
\end{array}%
\right) \text{ }.  \label{11}
\end{equation}%
Since the spinors are real the conjugate spinors are $\bar{\chi}=\tilde{\chi}%
\gamma _{0}$.

Without fermions the right hand sides in (\ref{9}) vanish and it is easy to
show that in this case equations (\ref{8}) and (\ref{9}) are equivalent to
the old integrable matrix equation $\eta ^{\mu \nu }(\alpha \mathbf{G}^{-1}%
\mathbf{G}_{,\nu })_{,\mu }=0$ considered in \cite{BZ1}.\ To see this one
need to use relations (\ref{2}) and (\ref{5}) from which follows:%
\begin{equation}
\mathbf{G}^{-1}\mathbf{G}_{,\mu }=2\mathbf{\tilde{V}}^{-1}\mathbf{P}_{\mu }%
\mathbf{\tilde{V}}\text{ }.  \label{12}
\end{equation}%
Then our old equation becomes $\eta ^{\mu \nu }\left[ \alpha ^{-1}(\alpha 
\mathbf{P}_{\mu })_{,\nu }+\mathbf{Q}_{\nu }\mathbf{P}_{\mu }-\mathbf{P}%
_{\mu }\mathbf{Q}_{\nu }\right] =0.$ Its trace gives (\ref{8}) and its other
components reproduce (\ref{9}) with zeros in the right hand sides.

In paper \cite{N2} it was claimed that the full supergravitational system (%
\ref{8})-(\ref{10}) can be integrated with the help of the following linear
spectral problem: 
\begin{gather}
\mathbf{\hat{V}}^{-1}\mathbf{\hat{V}}_{,\mu }=\mathbf{Q}_{\mu }+\frac{1+t^{2}%
}{1-t^{2}}\mathbf{P}_{\mu }+\frac{2t}{1-t^{2}}\varepsilon _{\mu \nu }\eta
^{\nu \lambda }\mathbf{P}_{\lambda }-  \label{13} \\
-\frac{2t^{2}}{\left( 1-t^{2}\right) ^{2}}3i\bar{\chi}^{\left( 1\right)
}\gamma _{\mu }\chi ^{\left( 2\right) }\mathbf{Y}^{3}-\frac{t\left(
1+t^{2}\right) }{\left( 1-t^{2}\right) ^{2}}3i\bar{\chi}^{\left( 1\right)
}\gamma _{0}\gamma _{1}\gamma _{\mu }\chi ^{\left( 2\right) }\mathbf{Y}^{3}.
\notag
\end{gather}%
Here $\mathbf{\hat{V}}=\mathbf{\hat{V}}\left( x^{0},x^{1},t\right) $ and the
quantity $t$ is the complex spectral parameter which depends on coordinates $%
x^{\mu }$ satisfying the differential equation: 
\begin{equation}
\frac{t_{,\mu }}{t}=\frac{1+t^{2}}{1-t^{2}}\frac{\alpha _{,\mu }}{\alpha }+%
\frac{2t}{1-t^{2}}\varepsilon _{\mu \nu }\eta ^{\nu \lambda }\frac{\alpha
_{,\lambda }}{\alpha }  \label{14}
\end{equation}%
[its self-consistency requirement $t_{,\mu \nu }=t_{,\nu \mu }$ is carried
out automatically due to the condition (\ref{8})]. Here $\varepsilon _{\mu
\nu }$ is the two-dimensional Euclidean antisymmetric symbol ($\varepsilon
_{01}=1$).

The solution of the equation (\ref{14}) contains one arbitrary complex
constant $w$ then parameter $t=t(x^{0},x^{1},w)$ has one arbitrary degree of
freedom in addition to changing of coordinates. But all terms $\mathbf{Q}%
_{\mu },$ $\mathbf{P}_{\mu },$ $\chi ^{\left( 1\right) },$ $\chi ^{\left(
2\right) }$ in the right hand side of (\ref{13}) are functions on the two
coordinates $x^{0},x^{1}$ only, that is they are treated as unknown
"potentials" independent on the spectral parameter $t$. The equations (\ref%
{8})-(\ref{10}) for these "potentials" should result from the linear
spectral system (\ref{13}) as its self-consistency conditions.

However, it is evident that self-consistency conditions of the system (\ref%
{13}) can not reproduce equations (\ref{10}) since the last two terms in
this system are quadratic in fermions. Consequently, all self-consistency
conditions which one can get from (\ref{13}) contain only products of
fermions of even powers. In graded algebra the knowledge of such products is
far to be enough to reconstruct their anticommuting multipliers.

Let's derive the complete set of the self-consistency conditions of the Lax
representation (\ref{13}). To do this we first adduce it to the simpler form
more convenient for calculations which form in its bosonic part is the same
as in paper \cite{BZ1} [this reduction will be appropriate also for the
further extension of the system (\ref{13})]. Instead of the original
spectral matrix $\mathbf{\hat{V}}\left( x^{0},x^{1},t\right) $ we introduce
new one $\mathbf{\hat{G}}\left( x^{0},x^{1},t\right) $ by the relation \ 
\begin{equation}
\mathbf{\hat{G}}\left( x^{0},x^{1},t\right) =\mathbf{\hat{V}}\left(
x^{0},x^{1},t\right) \mathbf{\tilde{V}}\left( x^{0},x^{1}\right) \text{ }
\label{15}
\end{equation}%
and instead of the parameter $t\left( x^{0},x^{1},w\right) $ we use the new
spectral parameter $s\left( x^{0},x^{1},w\right) $:%
\begin{equation}
s=\alpha t.  \label{16}
\end{equation}%
If we pass to the light-like coordinates $\zeta $ and $\eta $: 
\begin{equation}
\zeta =\frac{1}{\sqrt{2}}\left( x^{0}-x^{1}\right) \text{ },\text{ }\eta =%
\frac{1}{\sqrt{2}}\left( x^{0}+x^{1}\right) \text{ },  \label{17}
\end{equation}%
the system (\ref{13}) takes the following form: \ \ \ 
\begin{eqnarray}
\mathbf{\hat{G}}^{-1}\mathbf{\hat{G}}_{,\zeta } &=&\frac{\alpha }{\alpha -s}%
\mathbf{G}^{-1}\mathbf{G}_{,\zeta }+\frac{\alpha s}{\left( \alpha -s\right)
^{2}}\mathbf{G}^{-1}\mathbf{Y}^{3}u_{-}\text{ },  \label{18} \\
\mathbf{\hat{G}}^{-1}\mathbf{\hat{G}}_{,\eta } &=&\frac{\alpha }{\alpha +s}%
\mathbf{G}^{-1}\mathbf{G}_{,\eta }-\frac{\alpha s}{\left( \alpha +s\right)
^{2}}\mathbf{G}^{-1}\mathbf{Y}^{3}u_{+}\text{ },  \notag
\end{eqnarray}%
where%
\begin{eqnarray}
u_{-} &=&-\frac{3i\alpha }{\sqrt{2}}\bar{\chi}^{\left( 1\right) }\left(
\gamma _{0}-\gamma _{1}\right) \chi ^{\left( 2\right) },  \label{19} \\
u_{+} &=&-\frac{3i\alpha }{\sqrt{2}}\bar{\chi}^{\left( 1\right) }\left(
\gamma _{0}+\gamma _{1}\right) \chi ^{\left( 2\right) }.  \notag
\end{eqnarray}%
In (\ref{18}) $\mathbf{\hat{G}=\hat{G}}\left( \zeta ,\eta ,s\right) $ while $%
\mathbf{G},$\ $u_{-},$ $u_{+}$ are independent on the spectral parameter $s$%
. From (\ref{14}) and (\ref{16}) follows that $s$ satisfies the
requirements: 
\begin{equation}
\frac{s_{,\zeta }}{s}=\frac{2\alpha _{,\zeta }}{\alpha -s}\text{ },\text{ }%
\frac{s_{,\eta }}{s}=\frac{2\alpha _{,\eta }}{\alpha +s}\text{ }.  \label{20}
\end{equation}%
Now it is easy to get the self-consistency conditions of (\ref{18}) from the
identity $\mathbf{\hat{G}}_{,\zeta \eta }-\mathbf{\hat{G}}_{,\eta \zeta }=0$%
. After substitution into it the second derivatives of $\mathbf{\hat{G}}$
[they follow from the same equations (\ref{18})] and multiplying it by $%
\left( s+\alpha \right) ^{2}\left( s-\alpha \right) ^{2}$ we obtain the
cubic polynomial in $s$. Demanding the coefficients of this polynomial to be
zero we get the following three equations (the fourth one is satisfied
automatically after these three are fulfilled):

\begin{equation}
\left( \alpha \mathbf{G}^{-1}\mathbf{G}_{,\zeta }-\alpha \mathbf{G}^{-1}%
\mathbf{Y}^{3}u_{-}\right) _{,\eta }+\left( \alpha \mathbf{G}^{-1}\mathbf{G}%
_{,\eta }-\alpha \mathbf{G}^{-1}\mathbf{Y}^{3}u_{+}\right) _{,\zeta }=0\text{
},  \label{22}
\end{equation}%
\begin{equation}
\left( u_{+}\right) _{,\zeta }=0\text{ },\text{ }\left( u_{-}\right) _{,\eta
}=0.  \label{23}
\end{equation}%
This system indeed contain the scalar equations (\ref{9}). To show this it
is necessary to return to the coordinates $x^{0},x^{1}$ and pass to the
variables $\mathbf{V,Q}_{\mu },\mathbf{P}_{\mu }$ instead of $\mathbf{G}$
and $\mathbf{G}^{-1}\mathbf{G}_{,\mu }$. In course of these calculations one
need to use relations (\ref{2}), (\ref{4})-(\ref{7}) and identity%
\begin{equation}
\mathbf{\tilde{V}}^{-1}\mathbf{Y}^{3}\mathbf{\tilde{V}=}\alpha \mathbf{G}%
^{-1}\mathbf{Y}^{3}.  \label{23-1}
\end{equation}

So then the solution of spectral problem (\ref{13}) permit to find the
metric $\mathbf{G}$ and two components of "torsion" $\bar{\chi}^{\left(
1\right) }\gamma _{\mu }\chi ^{\left( 2\right) }$ but not the fermions $\chi
^{\left( 1\right) },\chi ^{\left( 2\right) }$ itself, that is this system do
not cover the full fermionic equations (\ref{10}). Of course, the partial
result (\ref{23}) should be compatible with equations (\ref{10}) and, as we
show later, this is indeed the case\footnote{%
The surprisingly trivial equations (\ref{23}) for the "torsion" components
are interesting and new. It seems that this result has not been indicated in
the previous literature.}.

\section{The complete spectral linear system}

To construct an extension of the spectral representation (\ref{18})-(\ref{20}%
) covering all equations of motion (\ref{8})-(\ref{10}) it is necessary to
add to $\mathbf{\hat{G}}\left( \zeta ,\eta ,s\right) $ some fermionic
matrices depending on the variables $\zeta ,\eta ,s$ and to find an
appropriate spectral representation for the such composite set. The simplest
way to do this is to use the 4-dimensional superspace parametrized by the
coordinates $(\zeta ,\eta ,\theta _{1},\theta _{2})$ with two odd elements $%
\theta _{1}$ and $\theta _{2}$ and consider bosonic matrix $\mathbf{\hat{G}}%
\left( \zeta ,\eta ,s\right) $ (with even entries) and additional fermionic
matrices $\mathbf{\hat{\Omega}}_{1}\left( \zeta ,\eta ,s\right) $ and $%
\mathbf{\hat{\Omega}}_{2}\left( \zeta ,\eta ,s\right) $ (with odd entries)
as components of the single spectral supermatrix $\mathbf{\hat{\Psi}(}\zeta
,\eta ,\theta _{1},\theta _{2},s)$ (see the analogous technique for the
chiral fields in two-dimensional Minkowski space-time in \cite{Mik}):%
\begin{equation}
\mathbf{\hat{\Psi}}=\mathbf{\hat{G}}\left( \mathbf{I+}\text{ }\theta _{1}%
\mathbf{\hat{\Omega}}_{1}+\theta _{2}\mathbf{\hat{\Omega}}_{2}+\theta
_{1}\theta _{2}\mathbf{\hat{H}}\text{ }\right) ,  \label{24}
\end{equation}%
where $\mathbf{\hat{H}}\left( \zeta ,\eta ,s\right) $ (with even entries) is
a standard auxiliary addend. In superspace the generalization $\mathbf{\Psi }%
\mathbb{(}\zeta ,\eta ,\theta _{1},\theta _{2})$ of the matrix $\mathbf{G(}%
\zeta ,\eta )$ can be written as 
\begin{equation}
\mathbf{\Psi }\text{ }\mathbf{=G}\left( \mathbf{I+}\theta _{1}\mathbf{\Omega 
}_{1}+\theta _{2}\mathbf{\Omega }_{2}+\theta _{1}\theta _{2}\mathbf{H}\right)
\label{25}
\end{equation}%
where $\mathbf{\Omega }_{1}(\zeta ,\eta )$ and $\mathbf{\Omega }_{2}(\zeta
,\eta )$ consist of the odd entries and auxiliary matrix $\mathbf{H}(\zeta
,\eta )$ has even components. The supermarix $\mathbf{\Psi }$, as well as
its progenitor $\mathbf{G,}$ should be symmetric: \ 
\begin{equation}
\mathbf{\tilde{\Psi}}\mathbb{=}\mathbf{\Psi }\text{ }.  \label{25-1}
\end{equation}%
Going after the machinery of supersymmetry we introduce two
super-differential operators%
\begin{equation}
D_{\zeta }=\frac{\partial }{\partial \theta _{2}}-\theta _{2}\frac{\partial 
}{\partial \zeta }\text{ },\text{ }D_{\eta }=-\frac{\partial }{\partial
\theta _{1}}+\theta _{1}\frac{\partial }{\partial \eta }\text{ },  \label{28}
\end{equation}%
which anticommute with each other: 
\begin{equation}
D_{\zeta }D_{\eta }\mathfrak{+}D_{\eta }D_{\zeta }=0\text{ }  \label{29}
\end{equation}%
and consider the following superspace Lax representation for the fields (\ref%
{24}) and (\ref{25}): 
\begin{equation}
\mathbf{\hat{\Psi}}^{-1}D_{\zeta }\mathbf{\hat{\Psi}}\text{ }\mathbf{=}\frac{%
\alpha }{\alpha -s}\mathbf{\Psi }^{-1}D_{\zeta }\mathbf{\Psi }\text{ }%
\mathbf{,}\text{ \ }\mathbf{\hat{\Psi}}^{-1}D_{\eta }\mathbf{\hat{\Psi}}%
\text{ }\mathbf{=}\frac{\alpha }{\alpha +s}\mathbf{\Psi }^{-1}D_{\eta }%
\mathbf{\Psi }\text{\ }.  \label{30}
\end{equation}%
By the direct calculations it can be shown that the only condition of
self-consistency (that is of the requirement $D_{\zeta }D_{\eta }\mathbf{%
\hat{\Psi}}\mathfrak{+}D_{\eta }D_{\zeta }\mathbf{\hat{\Psi}}=0$) for this
superspace spectral problem is: 
\begin{equation}
D_{\zeta }\left[ \alpha \mathbf{\Psi }^{-1}D_{\eta }\mathbf{\Psi }\right]
-D_{\eta }\left[ \alpha \mathbf{\Psi }^{-1}D_{\zeta }\mathbf{\Psi }\right] =0
\label{32}
\end{equation}%
Now we should insert into this equation the matrix $\mathbf{\Psi }$ from (%
\ref{25}) and equate to zero coefficients in front of $\theta _{1},$ $\theta
_{2},$ $\theta _{1}\theta _{2}$ and also the term independent on the odd
coordinates. This gives the following four equations: 
\begin{equation}
\left( \alpha \mathbf{G}^{-1}\mathbf{G}_{,\zeta }+\alpha \mathbf{\Omega }%
_{2}^{2}\right) _{,\eta }+\left( \alpha \mathbf{G}^{-1}\mathbf{G}_{,\eta
}+\alpha \mathbf{\Omega }_{1}^{2}\right) _{,\zeta }=0\text{ },  \label{33}
\end{equation}%
\begin{equation}
2\mathbf{\Omega }_{1,\zeta }+\frac{\alpha _{,\zeta }}{\alpha }\mathbf{\Omega 
}_{1}+\mathbf{G}^{-1}\mathbf{G}_{,\zeta }\mathbf{\Omega }_{1}-\mathbf{\Omega 
}_{1}\mathbf{G}^{-1}\mathbf{G}_{,\zeta }+\frac{1}{2}\left( \mathbf{\Omega }%
_{2}^{2}\mathbf{\Omega }_{1}-\mathbf{\Omega }_{1}\mathbf{\Omega }%
_{2}^{2}\right) =0\text{ },  \label{34}
\end{equation}%
\begin{equation}
2\mathbf{\Omega }_{2,\eta }+\frac{\alpha _{,\eta }}{\alpha }\mathbf{\Omega }%
_{2}+\mathbf{G}^{-1}\mathbf{G}_{,\eta }\mathbf{\Omega }_{2}-\mathbf{\Omega }%
_{2}\mathbf{G}^{-1}\mathbf{G}_{,\eta }+\frac{1}{2}\left( \mathbf{\Omega }%
_{1}^{2}\mathbf{\Omega }_{2}-\mathbf{\Omega }_{2}\mathbf{\Omega }%
_{1}^{2}\right) =0\text{ },  \label{35}
\end{equation}%
\begin{equation}
\mathbf{H=}\frac{1}{2}\left( \mathbf{\Omega }_{2}\mathbf{\Omega }_{1}-%
\mathbf{\Omega }_{1}\mathbf{\Omega }_{2}\right) \text{ }.  \label{36}
\end{equation}

We demonstrate below that spinors of interest $\chi ^{\left( 1\right)
}\left( \zeta ,\eta \right) $ and $\chi ^{\left( 2\right) }\left( \zeta
,\eta \right) $ are encoded in the odd entries of the fermionic matrices $%
\mathbf{\Omega }_{1}$ and $\mathbf{\Omega }_{2}$ and equations (\ref{34})
and (\ref{35}) are the same equations (\ref{10}) missed in the original
spectral representation (\ref{13}).

From (\ref{34})-(\ref{35}) follows two relations containing only squares $%
\mathbf{\Omega }_{1}^{2}$ and $\mathbf{\Omega }_{2}^{2}$ of the odd matrices 
$\mathbf{\Omega }_{1}$ and $\mathbf{\Omega }_{2}.$ To see this one should
multiply (\ref{34}) first from the right and then from the left by $\mathbf{%
\Omega }_{1}$ and take the sum of the results. The same trick should be
applied to (\ref{35}) but multiplying by $\mathbf{\Omega }_{2}.$ The outcome
of these operations is\footnote{%
In general in equations (\ref{37})-(\ref{38}) should appear also the
commutator $\mathbf{\Omega }_{1}^{2}\mathbf{\Omega }_{2}^{2}-\mathbf{\Omega }%
_{2}^{2}\mathbf{\Omega }_{1}^{2}$ but in our case it vanish. This happens
because we need to provide the super-metric $\mathbf{\Psi }$ to be
symmetric. If so then also the components $\mathbf{G\Omega }_{1}$ and $%
\mathbf{G\Omega }_{2}$ are symmetric and it is easy to prove that due to the
symmetry of $\mathbf{G}$ and oddness of $\mathbf{\Omega }_{1}$ and $\mathbf{%
\Omega }_{2}$\textbf{\ }the\textbf{\ }aforementioned commutator is zero.%
\textbf{\ }}:%
\begin{equation}
\frac{1}{\alpha }\left( \alpha \mathbf{\Omega }_{1}^{2}\right) _{,\zeta }+%
\frac{1}{2}\left( \mathbf{G}^{-1}\mathbf{G}_{,\zeta }\mathbf{\Omega }%
_{1}^{2}-\mathbf{\Omega }_{1}^{2}\mathbf{G}^{-1}\mathbf{G}_{,\zeta }\right)
=0\text{ },  \label{37}
\end{equation}%
\begin{equation}
\frac{1}{\alpha }\left( \alpha \mathbf{\Omega }_{2}^{2}\right) _{,\eta }+%
\frac{1}{2}\left( \mathbf{G}^{-1}\mathbf{G}_{,\eta }\mathbf{\Omega }_{2}^{2}-%
\mathbf{\Omega }_{2}^{2}\mathbf{G}^{-1}\mathbf{G}_{,\eta }\right) =0\text{ }.
\label{38}
\end{equation}%
Then it is easy to show that these two bosonic relations together with
equation (\ref{33}) are equivalent to the system (\ref{22}) and (\ref{23})
that is namely to that one which indeed is covered by the original spectral
representation (\ref{13}). This result proves the statement which we
formulated earlier: the equations (\ref{23}) for the "torsion" are
compatible with fermionic equations (\ref{10}).

Another useful result is coming from the comparison of equations (\ref{33})
and (\ref{22}). It gives the connection between squares of the $\mathbf{%
\Omega }$-matrices and quadratic combinations of spinors:

\begin{eqnarray}
\mathbf{\Omega }_{1}^{2} &=&-\mathbf{G}^{-1}\mathbf{Y}^{3}u_{+}=\frac{3i}{%
\sqrt{2}}\bar{\chi}^{\left( 1\right) }\left( \gamma _{0}+\gamma _{1}\right)
\chi ^{\left( 2\right) }\alpha \mathbf{G}^{-1}\mathbf{Y}^{3}\text{ },
\label{39} \\
\mathbf{\Omega }_{2}^{2} &=&-\mathbf{G}^{-1}\mathbf{Y}^{3}u_{-}=\frac{3i}{%
\sqrt{2}}\bar{\chi}^{\left( 1\right) }\left( \gamma _{0}-\gamma _{1}\right)
\chi ^{\left( 2\right) }\alpha \mathbf{G}^{-1}\mathbf{Y}^{3}\text{ }.  \notag
\end{eqnarray}%
In the moment we will use this result to find the way how to express the
spinors $\chi ^{\left( 1\right) }$ and $\chi ^{\left( 2\right) }$ itself in
terms of the $\mathbf{\Omega }$-matrices. This task is important part of our
approach because the inverse scattering method being applied to the spectral
problem (\ref{30}) gives solution namely for the matrices $\mathbf{\Omega }%
_{1}$ and $\mathbf{\Omega }_{2}$ and not directly for spinors.

We see that the super-Lax pair (\ref{30}) apart of the property that it is
complete [that is covering the whole system (\ref{8})-(\ref{10}) of the
equations of motion] also has no second order poles with respect to the
spectral parameter which property permits to use the standard technique to
extract the exact solitonic solutions from this spectral problem.

It is easy to understand the source of the second order poles in
representation (\ref{13}) or (\ref{18}). Of course, the best approach to
integration is to work directly in superspace with the problem (\ref{30})
containing only simple poles and in the last three sections of the present
paper we will outline this way. However, one could prefer to begin with by
splitting representation (\ref{30}) into component equations for the
matrices $\mathbf{\hat{G},}$ $\mathbf{\hat{\Omega}}_{1},$ $\mathbf{\hat{%
\Omega}}_{2},$ $\mathbf{\hat{H}.}$ This mode technically is more cumbersome
but also possible. In doing so one should insert expressions (\ref{24})-(\ref%
{25}) for the matrices $\mathbf{\hat{\Psi}}$ and $\mathbf{\Psi }$ into
equation (\ref{30}) and equate to zero all matrix coefficients in the
resulting $\theta $-polynomial. From the free term and terms linear in $%
\theta _{1}$ and $\theta _{2}$ follow (among other relations) equations with
second order poles in spectral parameter $s$ [these are nothing else but
equations (\ref{18}) in which one should replace the terms $\mathbf{G}^{-1}%
\mathbf{Y}^{3}u_{+}$ and $\mathbf{G}^{-1}\mathbf{Y}^{3}u_{-}$ by minus $%
\mathbf{\Omega }_{1}^{2}$ and minus $\mathbf{\Omega }_{2}^{2}$ respectively,
see (\ref{39})]. This is explanation of the supergravitational effect of the
appearance of the second order poles revealed in papers \cite{N1, N2}.

Let's show now that fermionic equations (\ref{34})-(\ref{35}) for matrices $%
\mathbf{\Omega }_{1}$ and $\mathbf{\Omega }_{2}$ are the same that original
equations (\ref{10}) for spinors $\chi ^{\left( 1\right) }$ and $\chi
^{\left( 2\right) }$ and how one can extract these spinors from known
solution for $\mathbf{\Omega }_{1}$ and $\mathbf{\Omega }_{2}.$ The spinors
are:%
\begin{equation}
\chi ^{\left( 1\right) }=\binom{a_{1}}{b_{1}}\text{ },\text{ }\chi ^{\left(
2\right) }=\binom{a_{2}}{b_{2}}\text{ },  \label{40}
\end{equation}%
where all components $a,b$ are odd elements. Then%
\begin{eqnarray}
\bar{\chi}^{\left( 1\right) }\left( \gamma _{0}+\gamma _{1}\right) \chi
^{\left( 2\right) } &=&\left( a_{1}+b_{1}\right) \left( a_{2}+b_{2}\right) 
\text{ },  \label{41} \\
\bar{\chi}^{\left( 1\right) }\left( \gamma _{0}-\gamma _{1}\right) \chi
^{\left( 2\right) } &=&\left( a_{1}-b_{1}\right) \left( a_{2}-b_{2}\right) 
\text{ }.  \notag
\end{eqnarray}%
From (\ref{39}), taking into account the identity (\ref{23-1}), follows:%
\begin{eqnarray}
\mathbf{\Omega }_{1}^{2} &=&\frac{3i}{\sqrt{2}}\mathbf{\tilde{V}}^{-1}\left[
\left( a_{1}+b_{1}\right) \left( a_{2}+b_{2}\right) \mathbf{Y}^{3}\right] 
\mathbf{\tilde{V}},  \label{42} \\
\mathbf{\Omega }_{2}^{2} &=&\frac{3i}{\sqrt{2}}\mathbf{\tilde{V}}^{-1}\left[
\left( a_{1}-b_{1}\right) \left( a_{2}-b_{2}\right) \mathbf{Y}^{3}\right] 
\mathbf{\tilde{V}.}  \notag
\end{eqnarray}%
It is clear that matrices $\mathbf{\Omega }_{1}$ and $\mathbf{\Omega }_{2}$
should be some linear combinations (with even matrix coefficients) of the
quantities $a,b$ and these combinations should satisfy the requirements (\ref%
{42}). Simple consideration lead to the result that we can chose $\mathbf{%
\Omega }_{1}$ and $\mathbf{\Omega }_{2}$ as:%
\begin{eqnarray}
\mathbf{\Omega }_{1} &=&\left( 3i\right) ^{1/2}2^{-3/4}\mathbf{\tilde{V}}%
^{-1}\left[ \left( a_{1}+b_{1}\right) \mathbf{Y}^{1}+\left(
a_{2}+b_{2}\right) \mathbf{Y}^{2}\right] \mathbf{\tilde{V}}\text{ },
\label{43} \\
\mathbf{\Omega }_{2} &=&\left( 3i\right) ^{1/2}2^{-3/4}\mathbf{\tilde{V}}%
^{-1}\left[ \left( a_{1}-b_{1}\right) \mathbf{Y}^{1}+\left(
a_{2}-b_{2}\right) \mathbf{Y}^{2}\right] \mathbf{\tilde{V}}\text{ },  \notag
\end{eqnarray}%
which expressions (due to the identity $\mathbf{Y}^{1}\mathbf{Y}^{2}-\mathbf{%
Y}^{2}\mathbf{Y}^{1}=2\mathbf{Y}^{3}$) automatically meet the requirements (%
\ref{42}). Now the little bit long calculations show that equations (\ref{34}%
)-(\ref{35}) being transformed to the coordinates $x^{0},x^{1}$ and after
substitution $\mathbf{\Omega }_{1}$ and $\mathbf{\Omega }_{2}$ in the form (%
\ref{43}) coincide with the original fermionic equations (\ref{10}).

From relations (\ref{43}) can be found all four components of spinors $\chi
^{\left( 1\right) }$ and $\chi ^{\left( 2\right) }$: 
\begin{eqnarray}
a_{1}\mathbf{Y}^{1}+a_{2}\mathbf{Y}^{2} &=&\left( 3i\right)
^{-1/2}2^{-1/4}\left( \mathbf{\tilde{V}\Omega }_{1}\mathbf{\tilde{V}}^{-1}+%
\mathbf{\tilde{V}\Omega }_{2}\mathbf{\tilde{V}}^{-1}\right) ,  \label{43-1}
\\
b_{1}\mathbf{Y}^{1}+b_{2}\mathbf{Y}^{2} &=&\left( 3i\right)
^{-1/2}2^{-1/4}\left( \mathbf{\tilde{V}\Omega }_{1}\mathbf{\tilde{V}}^{-1}-%
\mathbf{\tilde{V}\Omega }_{2}\mathbf{\tilde{V}}^{-1}\right) ,  \notag
\end{eqnarray}%
if matrices $\mathbf{\tilde{V}\Omega }_{1}\mathbf{\tilde{V}}^{-1}$ and $%
\mathbf{\tilde{V}\Omega }_{2}\mathbf{\tilde{V}}^{-1}$ are known. These
matrices, as can be seen from the last formulas, should be symmetric and
traceless. First of these requirements is the same as symmetry condition (%
\ref{25-1}), where from follows that also $\mathbf{G\Omega }_{1}$ and $%
\mathbf{G\Omega }_{2}$ (that is $\mathbf{V\tilde{V}\Omega }_{1}$ and $%
\mathbf{V\tilde{V}\Omega }_{2})$ are symmetric. The traceless conditions 
\begin{equation}
\text{Tr}\mathbf{\Omega }_{1}=0\text{ },\text{ Tr}\mathbf{\Omega }_{2}=0
\label{44}
\end{equation}%
can be easily ensured as simple additional constrains to the equations (\ref%
{34})-(\ref{35}) because traces of these equations are $\left( \sqrt{\alpha }%
\text{Tr}\mathbf{\Omega }_{1}\right) _{,\zeta }=0$ and $\left( \sqrt{\alpha }%
\text{Tr}\mathbf{\Omega }_{2}\right) _{,\eta }$ $=0.$

Now we came to the main problem: how to find solutions for matrices $\mathbf{%
G,\Omega }_{1,}\mathbf{\Omega }_{2}$. Unfortunately there are no miracles
and in general this is impossible. However, for the special case of
solitonic type of these fields the exact solutions can be constructed and we
show this in the next section.

\section{Super-solitonic solutions of the spectral problem}

From the basic equations (\ref{30}) can be seen that the supermatrix of
interest $\mathbf{\Psi }$ is equal to $\mathbf{\hat{\Psi}}$ at zero value of
the spectral parameter $s$:%
\begin{equation}
\mathbf{\hat{\Psi}}_{s=0}=\mathbf{\Psi }\mathfrak{.}  \label{45}
\end{equation}%
However, we need solution for $\mathbf{\Psi }$ under some additional
conditions. All necessary reality conditions as well as the requirements (%
\ref{44}) can be ensured at the end of calculations by the appropriate
choice of the arbitrary functions and constants which will appear in the
course of integration of equations (\ref{30}). The requirement (\ref{4})
which is the same as $\det \mathbf{G=}\alpha ^{2}$ can be satisfied by the
simple renormalization of the solution in the way used for the same purpose
in paper \cite{BZ1}. If at the end of calculations we will obtain solution
with $\det \mathbf{G\neq }\alpha ^{2}$ then we can pass to the new solution
corresponding to the new matrix $\mathbf{\acute{G}}=\alpha \left( \det 
\mathbf{G}\right) ^{-1/2}\mathbf{G}.$ It is easy to show that the system (%
\ref{33})-(\ref{36}) is invariant under this transformation but new matrix
automatically satisfies the necessary condition $\det \mathbf{\acute{G}=}%
\alpha ^{2}.$

More complicated task is to ensure the requirement (\ref{25-1}) of the
symmetry of the supermatrix $\mathbf{\Psi }\mathfrak{.}$ In principle also
the class of symmetric supermatrices $\mathbf{\Psi }$ can be singled out
from the general solution of spectral equations (\ref{30}) at the end of
calculations by an appropriate fixation of the arbitrariness which will be
present in the solution. However, technically this is difficult task. Much
more convenient way is to admit from the outset a suitable additional
restrictions on the solutions of equations (\ref{30}) which are compatible
with these equations and will guarantee the symmetricity of $\mathbf{\Psi }$
already from the first steps of the integration procedure. This approach was
used in \cite{BZ1} and it works well also here. To clarify this point we
remind that procedure of integration of the spectral equations (\ref{30})
comes from the so-called "dressing technique", that is it assumes that we
know some background solution $\alpha ,$ $\mathbf{\Psi }_{0}$ (with
symmetric $\mathbf{\Psi }_{0}$) of the equations (\ref{8}) and (\ref{32})
and corresponding solution $\mathbf{\hat{\Psi}}_{0}$ of equations (\ref{30})
which satisfy the relation $\left( \mathbf{\hat{\Psi}}_{0}\right) _{s=0}=%
\mathbf{\Psi }_{0}.$ Then any solution of the spectral equations for $%
\mathbf{\hat{\Psi}}$ can be represented in the form: 
\begin{equation}
\mathbf{\hat{\Psi}}=\mathbf{\hat{\Psi}}_{0}\mathbf{\hat{K}}.  \label{46}
\end{equation}%
The equations for the dressing supermatrix $\mathbf{\hat{K}}\left( \zeta
,\eta ,\theta _{1},\theta _{2},s\right) $ (which matrix is even) follows
from (\ref{30}) and they are:%
\begin{equation}
D_{\zeta }\mathbf{\hat{K}=}\frac{\alpha }{\alpha -s}\left[ \mathbf{\hat{K}%
\Psi }^{-1}\left( D_{\zeta }\mathbf{\Psi }\right) -\mathbf{\Psi }%
_{0}^{-1}\left( D_{\zeta }\mathbf{\Psi }_{0}\right) \mathbf{\hat{K}}\right] ,
\label{47}
\end{equation}%
\begin{equation}
D_{\eta }\mathbf{\hat{K}=}\frac{\alpha }{\alpha +s}\left[ \mathbf{\hat{K}%
\Psi }^{-1}\left( D_{\eta }\mathbf{\Psi }\right) \mathbf{-\Psi }%
_{0}^{-1}\left( D_{\eta }\mathbf{\Psi }_{0}\right) \mathbf{\hat{K}}\right] .
\label{48}
\end{equation}%
Assume that some $\mathbf{\hat{K}}\left( s\right) $ (for simplicity we do
not show arguments $\zeta ,\eta ,\theta _{1},\theta _{2}$ in all functions)
satisfies these equations. Replacing in it the argument $s$ by $\alpha
^{2}/s $ we construct the new supermatrix $\mathbf{\hat{K}}_{new}\left(
s\right) =\mathbf{\Psi }_{0}^{-1}\left[ \mathbf{\hat{K}}^{-1}\left( \alpha
^{2}/s\right) \right] _{trans}\mathbf{\Psi }$, where [and in the next
formula (\ref{49})] the index "trans" means that one should take
transposition of the matrix in square bracket. Now by the direct
calculations can be shown that $\mathbf{\hat{K}}_{new}$ also satisfies the
equations (\ref{47})-(\ref{48}) if $\mathbf{\Psi }$ is symmetric. We demand $%
\mathbf{\hat{K}}_{new}=\mathbf{\hat{K}}$ which guarantees the symmetry of
the supermatrix $\mathbf{\Psi .}$ It can be proved that this restriction is
the unique way to provide the necessary symmetry requirement. Thus, the
condition ensuring the symmetry of $\mathbf{\Psi }$ takes the form:%
\begin{equation}
\mathbf{\Psi }=\left[ \mathbf{\hat{K}}\left( s\right) \right] _{trans}%
\mathbf{\Psi }_{0}\mathbf{\hat{K}}\left( \alpha ^{2}/s\right) \text{ }.
\label{49}
\end{equation}%
To this relation we should add also the following boundary condition for $%
\mathbf{\hat{K}}$ at infinity of the complex plane of the spectral parameter:%
\begin{equation}
\mathbf{\hat{K}}_{s=\infty }=\mathbf{I.}  \label{50}
\end{equation}%
Under this boundary condition from (\ref{49}) we obtain$\mathfrak{\ }\mathbf{%
\Psi }$ $\mathbf{=\Psi }_{0}\left[ \mathbf{\hat{K}}\left( s\right) \right]
_{s=0}$ , that is the same result which follows from (\ref{45}) and (\ref{46}%
).

It is known \cite{BZ1} that solitonic solutions in pure gravity correspond
to the simple poles of the dressing matrix in the complex plane of $s.$ It
turns out that the same is true also in supergravity, i.e. for supersolitons
the dressing supermatrix has the simple meromorphic structure with respect
to the spectral parameter:%
\begin{equation}
\mathbf{\hat{K}}\text{ }\mathbf{=I+}\sum\limits_{k=1}^{\mathcal{N}}\text{ }%
\frac{\mathbf{R}_{k}\left( \zeta ,\eta ,\theta _{1},\theta _{2}\right) }{%
s-\mu _{k}\left( \zeta ,\eta \right) }\text{\ },  \label{51}
\end{equation}%
where $\mathbf{R}_{k}$ and $\mu _{k}$ are the even objects. Here and in the
sequel in this section six indices $k,l,m,n,p,q$ enumerate supersolitons
(the number of which is $\mathcal{N}$) and take values $1,2,...,\mathcal{N}.$
\textit{In this section no sums are assumed in expressions where these
indices are repeating, each time the summations over them is indicated
explicitly by the usual summation symbol.} The $\mathbf{\hat{K}}$ from (\ref%
{51}) we substitute to the equations (\ref{47})-(\ref{49}) from which follow
exact and unique solutions for the pole trajectories $\mu _{k}\left( \zeta
,\eta \right) $ and for the supermatrices $\mathbf{R}_{k}\left( \zeta ,\eta
,\theta _{1},\theta _{2}\right) $ in terms of the known solution for $%
\mathbf{\Psi }_{0}\left( \zeta ,\eta ,\theta _{1},\theta _{2}\right) .$
These calculations are almost literally the same as in the paper \cite{BZ1}
and we will not repeat them here (the only new point is that now we are in
the graded algebra and more care should be paid for disposition of the
different multipliers in course of the algebraic manipulations). The pole
trajectories $\mu _{k}$ are solutions of the following first order
differential equations:%
\begin{equation}
\mu _{k},_{\zeta }=\frac{2\alpha _{,\zeta }\mu _{k}}{\alpha -\mu _{k}},\text{
\ }\mu _{k},_{\eta }=\frac{2\alpha _{,\eta }\mu _{k}}{\alpha +\mu _{k}},
\label{52}
\end{equation}%
the solutions of which for each $k$ are the roots of the quadratic equation:%
\begin{equation}
\mu _{k}^{2}+2\left( \beta -w_{k}\right) \mu _{k}+\alpha ^{2}=0.  \label{53}
\end{equation}%
In this equation $w_{k}$ are the arbitrary complex constants, representing
the arbitrary constants of integration of the equations (\ref{52}), and $%
\beta \left( \zeta ,\eta \right) $ satisfy the same wave equation as $\alpha
\left( \zeta ,\eta \right) $ but it has to be chosen as second independent
solution of this equation, that is we take: 
\begin{equation}
\alpha =u\left( \zeta \right) +v\left( \eta \right) ,\text{ \ }\beta
=u\left( \zeta \right) -v\left( \eta \right) ,  \label{54}
\end{equation}%
where $u\left( \zeta \right) $ and $v\left( \eta \right) $ are two arbitrary
functions. As for the supermatrices $\mathbf{R}_{k\text{ }}$their components
are:%
\begin{equation}
\left( \mathbf{R}_{k}\right) ^{L}{}_{Q}=m_{\left( k\right)
}^{L}n_{Q}^{\left( k\right) }.  \label{55}
\end{equation}%
The supervectors $m_{\left( k\right) }^{L}\left( \zeta ,\eta ,\theta
_{1},\theta _{2}\right) $ and $n_{Q}^{\left( k\right) }\left( \zeta ,\eta
,\theta _{1},\theta _{2}\right) $ should be even. The supervectors $%
m_{\left( k\right) }^{L}$ follows from the direct integration of the
differential equations (\ref{47})-(\ref{48}) and the result of this
integration can be expressed in terms of the known background supermatrix $%
\mathbf{\hat{\Psi}}_{0}\left( \zeta ,\eta ,\theta _{1},\theta _{2},s\right) $
taken at $s=\mu _{k}$ and the set of the arbitrary constants $C_{L}^{\left(
k\right) }$ (which should be chosen to be even). These expressions are:%
\begin{equation}
m_{\left( k\right) }^{Q}\left( \zeta ,\eta ,\theta _{1},\theta _{2}\right) = 
\left[ \mathbf{\hat{\Psi}}_{0}^{-1}\left( \zeta ,\eta ,\theta _{1},\theta
_{2},s=\mu _{k}\right) \right] ^{QL}C_{L}^{\left( k\right) }.  \label{56}
\end{equation}%
After that the supervectors $n_{Q}^{\left( k\right) }\left( \zeta ,\eta
,\theta _{1},\theta _{2}\right) $ can be obtained from the additional
symmetry condition (\ref{49}) which means that they are solutions of the
following $\mathcal{N}$-order linear algebraic system: 
\begin{equation}
\sum_{l=1}^{\mathcal{N}}\frac{m_{\left( k\right) }^{Q}\left( \mathbf{\Psi }%
_{0}\right) _{QL}m_{\left( l\right) }^{L}}{\mu _{k}\mu _{l}-\alpha ^{2}}%
n_{J}^{\left( l\right) }=\mu _{k}^{-1}m_{\left( k\right) }^{L}\left( \mathbf{%
\Psi }_{0}\right) _{LJ}\text{ }.  \label{57}
\end{equation}%
To resolve this algebraic problem we need to calculate the $\mathcal{N}%
\times \mathcal{N}$ supermatrix (with respect to the indices $k$ and $l$)
which is inverse to the matrix $\mathbf{X}\left( \zeta ,\eta ,\theta
_{1},\theta _{2}\right) $ with components $(\mathbf{X)}_{kl}$: \ 
\begin{equation}
(\mathbf{X)}_{kl}=\frac{m_{\left( k\right) }^{Q}\left( \mathbf{\Psi }%
_{0}\right) _{QL}m_{\left( l\right) }^{L}}{\mu _{k}\mu _{l}-\alpha ^{2}}.
\label{58}
\end{equation}%
The components $(\mathbf{X)}_{kl}$ are symmetric in indices $k,l$ because
the background solution $\left( \mathbf{\Psi }_{0}\right) _{QL}$ is
symmetric in indices $Q,L$ and supervectors $m_{\left( k\right) }^{Q}$ are
even. The general form of $(\mathbf{X)}_{kl}$ is: 
\begin{equation}
(\mathbf{X)}_{kl}=A_{kl}+\theta _{1}B_{kl}+\theta _{2}C_{kl}+\theta
_{1}\theta _{2}D_{kl}\text{ },  \label{59}
\end{equation}%
where all components $A_{kl}\left( \zeta ,\eta \right) ,B_{kl}\left( \zeta
,\eta \right) ,C_{kl}\left( \zeta ,\eta \right) ,D_{kl}\left( \zeta ,\eta
\right) $ are symmetric in indices $k,l$ and are known since they can be
calculated explicitly from (\ref{58}) and (\ref{56}) in terms of the
background solutions $\mathbf{\hat{\Psi}}_{0},\mathbf{\Psi }_{0},\mu _{k}$
and constants $C_{L}^{\left( k\right) }.$ The components $A_{kl}$ and $%
D_{kl} $ are even, the components $B_{kl}$ and $C_{kl}$ are odd. The
calculations shows that the components $(\mathbf{X}^{-1})^{lk}$ of the
supermatrix which is inverse to the supermatrix with components $(\mathbf{X)}%
_{kl}$ [that is $\sum_{k=1}^{\mathcal{N}}(\mathbf{X}^{-1})^{lk}(\mathbf{X)}%
_{km}=\delta _{m}^{l}$] are:%
\begin{gather}
(\mathbf{X}^{-1})^{lk}=A^{lk}-\theta _{1}\sum_{m,n=1}^{\mathcal{N}%
}A^{lm}A^{kn}B_{mn}-\theta _{2}\sum_{m,n=1}^{\mathcal{N}}A^{lm}A^{kn}C_{mn}-
\label{60} \\
-\theta _{1}\theta _{2}\sum_{m,n=1}^{\mathcal{N}}A^{lm}A^{kn}\left[
D_{mn}+\sum_{p,q=1}^{\mathcal{N}}\left( B_{mp}C_{qn}+B_{np}C_{qm}\right)
A^{pq}\right] .  \notag
\end{gather}%
Here by $A^{kl}$ we designate the components of the matrix which is inverse
to the matrix with components $A_{kl}$ [that is $\sum_{k=1}^{\mathcal{N}%
}A^{lk}A_{km}=\delta _{m}^{l}$].

Now the solution of the algebraic system (\ref{57}) for supervectors $%
n_{Q}^{\left( k\right) }$ is: 
\begin{equation}
n_{Q}^{\left( k\right) }=\left( \sum_{l=1}^{\mathcal{N}}\mu _{l}^{-1}(%
\mathbf{X}^{-1})^{kl}m_{\left( l\right) }^{L}\right) \left( \mathbf{\Psi }%
_{0}\right) _{LQ}\text{ }.  \label{63}
\end{equation}%
Substituting this expression to (\ref{55}) we come to the solution for the
supermatrices $\mathbf{R}_{k}$:%
\begin{equation}
\left( \mathbf{R}_{k}\right) ^{L}{}_{Q}=m_{\left( k\right) }^{L}\left(
\sum_{l=1}^{\mathcal{N}}\mu _{l}^{-1}(\mathbf{X}^{-1})^{kl}m_{\left(
l\right) }^{M}\right) \left( \mathbf{\Psi }_{0}\right) _{MQ}\text{ },
\label{64}
\end{equation}%
where from, taking in account (\ref{51}) and relation $\mathbf{\Psi =\mathbf{%
\Psi }_{0}\hat{K}}_{s=0}$, follows the final $n$-solitonic solution for the
symmetric superfield $\mathbf{\Psi }$: 
\begin{equation}
\left( \mathbf{\Psi }\right) _{QL}=\left( \mathbf{\Psi }_{0}\right)
_{QL}-\sum_{k,l=1}^{\mathcal{N}}(\mathbf{X}^{-1})^{kl}L_{Q}^{\left( k\right)
}L_{L}^{\left( l\right) },  \label{65}
\end{equation}%
where the supervectors $L_{Q}^{\left( k\right) }$ are:%
\begin{equation}
L_{Q}^{\left( k\right) }=\mu _{k}^{-1}m_{\left( k\right) }^{L}\left( \mathbf{%
\Psi }_{0}\right) _{LQ}  \label{66}
\end{equation}

\section{Summary of prescriptions}

The previous section offer the detailed guide how to calculate the $\mathcal{%
N}$-solitonic solution for the matrix superfield $\mathbf{\Psi }\mathfrak{.}$
Let's summarize this program in the compressed set of the practical
prescriptions.

1. Chose some solutions $\alpha (\zeta ,\eta )$ of the wave equation (\ref{8}%
).

2. With this $\alpha (\zeta ,\eta )$ find some background solutions $\mathbf{%
G}^{\left( 0\right) }(\zeta ,\eta ),$ $\mathbf{\Omega }_{1}^{\left( 0\right)
}(\zeta ,\eta ),$ $\mathbf{\Omega }_{2}^{\left( 0\right) }(\zeta ,\eta ),$ $%
\mathbf{H}^{\left( 0\right) }(\zeta ,\eta )$ of the equations (\ref{33})-(%
\ref{36}) which satisfy the following additional conditions: $\det \mathbf{G}%
^{\left( 0\right) }=\alpha ^{2},$ Tr$\mathbf{\Omega }_{1}^{\left( 0\right)
}=0,$ Tr$\mathbf{\Omega }_{2}^{\left( 0\right) }=0$ and symmetricity of the
matrices $\mathbf{G}^{\left( 0\right) },$ $\mathbf{G}^{\left( 0\right) }%
\mathbf{\Omega }_{1}^{\left( 0\right) },$ $\mathbf{G}^{\left( 0\right) }%
\mathbf{\Omega }_{2}^{\left( 0\right) },$ $\mathbf{G}^{\left( 0\right) }%
\mathbf{H}^{\left( 0\right) }$. Substitute these solutions into the right
hand side of the relation (\ref{25}) which gives the symmetric seed
supermatrix $\mathbf{\Psi }_{0}(\zeta ,\eta ,\theta _{1},\theta _{2})$.

3. Chose some background frame $\mathbf{V}^{\left( 0\right) }(\zeta ,\eta ),$
satisfying requirement $\mathbf{G}^{\left( 0\right) }=\mathbf{V}^{\left(
0\right) }\mathbf{\tilde{V}}^{\left( 0\right) }$, and using solutions for $%
\mathbf{\Omega }_{1}^{\left( 0\right) }$ and $\mathbf{\Omega }_{2}^{\left(
0\right) }$find from formulas (\ref{43-1}) and (\ref{40}) the background
fermions $\chi _{\left( 0\right) }^{\left( 1\right) }(\zeta ,\eta )$ and $%
\chi _{\left( 0\right) }^{\left( 2\right) }(\zeta ,\eta )$.

4. Insert $\mathbf{\Psi }_{0}(\zeta ,\eta ,\theta _{1},\theta _{2})$ to the
right hand side of the Lax equations (\ref{30}) and integrate them to find
the background spectral supermatrix $\mathbf{\hat{\Psi}}_{0}(\zeta ,\eta
,\theta _{1},\theta _{2},s)$.

5. Chose the number $\mathcal{N}$ of solitons you wish to add to the
background and find the corresponding pole trajectories $\mu _{k}(\zeta
,\eta ,w_{k})$ as solutions of the quadratic equation (\ref{53}). The
quantities $w_{k}$ in this equations are arbitrary complex constants and
function $\beta (\zeta ,\eta )$ should be taken in accordance with
prescription (\ref{54}).

6. Find the inverse supermatrix $\left[ \mathbf{\hat{\Psi}}_{0}\right]
^{-1}(\zeta ,\eta ,\theta _{1},\theta _{2},s)$ and take it at the values of
spectral parameter $s=\mu _{k}$. This gives the supervectors $m_{\left(
k\right) }^{Q}\left( \zeta ,\eta ,\theta _{1},\theta _{2}\right) $ in
accordance with formula (\ref{56}).

7. Using the previous findings construct the $\mathcal{N}\times \mathcal{N}$
supermatrix $\mathbf{X}$ with entries $(\mathbf{X})_{kl}$ from the formulas (%
\ref{58})-(\ref{59}).

8. After that from (\ref{60}) calculate the inverse supermatrix $\mathbf{X}%
^{-1}$ with entries $(\mathbf{X}^{-1})^{lk}.$

9. Then from (\ref{65})-(\ref{66}) follows solution for the dressed matrix $%
\mathbf{\Psi }(\zeta ,\eta ,\theta _{1},\theta _{2})$ from which we obtain
the solutions under interest $\mathbf{G}(\zeta ,\eta ),$ $\mathbf{\Omega }%
_{1}(\zeta ,\eta ),$ $\mathbf{\Omega }_{2}(\zeta ,\eta ),$ $\mathbf{H}(\zeta
,\eta )$ using the expansion (\ref{25}) (check the requirements (\ref{44})
and conditions of symmetricity of the matrices $\mathbf{G},$ $\mathbf{%
G\Omega }_{1},$ $\mathbf{G\Omega }_{2},$ $\mathbf{GH}$).

10. Now construct any frame $\mathbf{V}(\zeta ,\eta )$ satisfying the
relation $\mathbf{G}=\mathbf{V\tilde{V}}$ and using previously founded
matrices $\mathbf{\Omega }_{1}$ and $\mathbf{\Omega }_{2}$ calculate the
components of spinors $\chi ^{\left( 1\right) }(\zeta ,\eta )$ and $\chi
^{\left( 2\right) }(\zeta ,\eta )$ from formulas (\ref{43-1}) and (\ref{40}%
). It is reasonable to fix the frame $\mathbf{V}$ by the same law it was
used for the fixation of the background frame $\mathbf{V}^{\left( 0\right) }$%
. The choice of the same gauge permits to see more clearly the structure of
the solitonic perturbations of the fermionic fields.

11. Finally provide all necessary reality conditions and renormalize
solution in order to satisfy the condition $\det \mathbf{G=}\alpha ^{2}.$

\section{Example of supersolitonic solution}

Here we are going to construct the concrete 1-supersoliton solution
following step by step the 11-point instruction given at the preceding
section just to demonstrate how these rules work. We will use one of the
simplest case of the diagonal background solution which produce a simple
structure of the bosonic soliton field $\mathbf{G}(\zeta ,\eta )$ and
corresponding solitonic fermions $\chi ^{\left( 1\right) }(\zeta ,\eta )$
and $\chi ^{\left( 2\right) }(\zeta ,\eta )$.

1. We choose some solution $\alpha (\zeta ,\eta )$ of the wave equation (\ref%
{8}).

2. Now we take the background solutions of the equations (\ref{33})-(\ref{36}%
) in the form of the diagonal matrices $\mathbf{G}^{\left( 0\right) },$ $%
\mathbf{\Omega }_{1}^{\left( 0\right) },$ $\mathbf{\Omega }_{2}^{\left(
0\right) },$ $\mathbf{H}^{\left( 0\right) }.$ One of the possibilities for
the matrix $\mathbf{G}^{\left( 0\right) }$ is the well known Kasner
solution: 
\begin{equation}
\mathbf{G}^{\left( 0\right) }=%
\begin{pmatrix}
\alpha ^{2p_{2}} & 0 \\ 
0 & \alpha ^{2p_{3}}%
\end{pmatrix}%
,\text{ \ }p_{2}+p_{3}=1\text{ },\text{ \ }\det \mathbf{G}^{\left( 0\right)
}=\alpha ^{2}\text{ },  \label{67}
\end{equation}%
where $p_{2}$ and $p_{3}$ are constants. This $\mathbf{G}^{\left( 0\right) }$
satisfies the equation (\ref{33}) because the squares of the diagonal odd
matrices $\Omega $ vanish identically. For the diagonal case the solutions
of the equations (\ref{34})-(\ref{35}) are trivial:%
\begin{equation}
\mathbf{\Omega }_{1}^{\left( 0\right) }=\alpha ^{-1/2}F_{1}(\eta )\mathbf{Y}%
^{1},\text{ \ }\mathbf{\Omega }_{2}^{\left( 0\right) }=\alpha
^{-1/2}F_{2}(\zeta )\mathbf{Y}^{1},  \label{68}
\end{equation}%
where $F_{1}(\eta )$ and $F_{2}(\zeta )$ are the arbitrary odd functions.
Then from the relation (\ref{36}) follows:%
\begin{equation}
\mathbf{H}^{\left( 0\right) }=\alpha ^{-1}F_{2}F_{1}\mathbf{I.}  \label{69}
\end{equation}%
The matrices $\mathbf{\Omega }_{1}^{\left( 0\right) }$ and \ $\mathbf{\Omega 
}_{2}^{\left( 0\right) }$ are traceless and all matrices $\mathbf{G}^{\left(
0\right) },$ $\mathbf{G}^{\left( 0\right) }\mathbf{\Omega }_{1}^{\left(
0\right) },$ $\mathbf{G}^{\left( 0\right) }\mathbf{\Omega }_{2}^{\left(
0\right) },$ $\mathbf{G}^{\left( 0\right) }\mathbf{H}^{\left( 0\right) }$
are symmetric as it should be. Substituting all of them into (\ref{25}) we
get the seed solution for the supermatrix $\mathbf{\Psi }_{0}$: 
\begin{equation}
\mathbf{\Psi }_{0}\text{ }\mathbf{=G}^{\left( 0\right) }\left( \mathbf{I+}%
\theta _{1}\mathbf{\Omega }_{1}^{\left( 0\right) }+\theta _{2}\mathbf{\Omega 
}_{2}^{\left( 0\right) }+\theta _{1}\theta _{2}\mathbf{H}^{\left( 0\right)
}\right)  \label{70}
\end{equation}

3. For the diagonal background $\mathbf{G}^{\left( 0\right) }$ (\ref{67})
the frame $\mathbf{V}^{\left( 0\right) }$ can be chosen also diagonal, that
is $\mathbf{V}^{\left( 0\right) }=daig(\alpha ^{p_{2}},\alpha ^{p_{3}})$. In
this frame formulas (\ref{43-1}) and (\ref{40}) give the following
background fermionic fields: 
\begin{equation}
\chi _{\left( 0\right) }^{\left( 1\right) }=\left( 3i\right)
^{-1/2}2^{-1/4}\alpha ^{-1/2}\binom{F_{1}(\eta )\mathbf{+}F_{2}(\zeta )}{%
F_{1}(\eta )\mathbf{-}F_{2}(\zeta )},\text{ \ }\chi _{\left( 0\right)
}^{\left( 2\right) }=0.  \label{71}
\end{equation}

4. Now we insert $\mathbf{\Psi }_{0}(\zeta ,\eta ,\theta _{1},\theta _{2})$
to the right hand side of the Lax equations (\ref{30}). The resulting system
is simple enough and can be integrated exactly. The solution for the
background spectral supermatrix $\mathbf{\hat{\Psi}}_{0}(\zeta ,\eta ,\theta
_{1},\theta _{2},s)$ is:%
\begin{gather}
\mathbf{\hat{\Psi}}_{0}=\mathbf{\hat{G}}^{\left( 0\right) }\left( \mathbf{I+}%
\theta _{1}\frac{\alpha }{\alpha +s}\mathbf{\Omega }_{1}^{\left( 0\right)
}+\theta _{2}\frac{\alpha }{\alpha -s}\mathbf{\Omega }_{2}^{\left( 0\right)
}\right) +  \label{72} \\
+\frac{1}{2}\theta _{1}\theta _{2}\mathbf{\hat{G}}^{\left( 0\right) }\left( 
\frac{\alpha }{\alpha -s}\mathbf{\Omega }_{2}^{\left( 0\right) }\mathbf{%
\Omega }_{1}^{\left( 0\right) }-\frac{\alpha }{\alpha +s}\mathbf{\Omega }%
_{1}^{\left( 0\right) }\mathbf{\Omega }_{2}^{\left( 0\right) }\right) \text{ 
},  \notag
\end{gather}%
where $\mathbf{\Omega }_{1}^{\left( 0\right) }$ and $\mathbf{\Omega }%
_{2}^{\left( 0\right) }$ are given by (\ref{68}) and 
\begin{equation}
\mathbf{\hat{G}}^{\left( 0\right) }(\zeta ,\eta ,s)=%
\begin{pmatrix}
\left( \alpha ^{2}+2\beta s+s^{2}\right) ^{p_{2}} & 0 \\ 
0 & \left( \alpha ^{2}+2\beta s+s^{2}\right) ^{p_{3}}%
\end{pmatrix}%
.  \label{73}
\end{equation}%
It is easy to check that this $\mathbf{\hat{\Psi}}_{0}$ at the spectral
point $s=0$ reduces to $\mathbf{\Psi }_{0}$ from (\ref{70}).

5. The number of solitons we wish to introduce on this background is just
one. Then, for simplicity, we can omit the index "1" in all expressions
where we enumerate the solitons. From quadratic equation (\ref{53}) follows
the pole trajectory $\mu (\zeta ,\eta ,w)$ of soliton. The quantity $w$ in
this equations is an arbitrary complex constants and function $\beta (\zeta
,\eta )$ should be taken in accordance with prescription (\ref{54}).
Consequently we have:%
\begin{equation}
\mu =w-\beta \pm \left[ \left( w-\beta \right) ^{2}-\alpha ^{2}\right]
^{1/2},  \label{74}
\end{equation}%
and both signs are acceptable.

6. The next step is to build the inverse supermatrix $\left( \mathbf{\hat{%
\Psi}}_{0}\right) ^{-1}$ and take it at the spectral point $s=\mu $. The
result is:%
\begin{eqnarray}
\left[ \left( \mathbf{\hat{\Psi}}_{0}\right) ^{-1}\right] _{s=\mu }
&=&\left( \mathbf{I-}\theta _{1}\frac{\sqrt{\alpha }F_{1}}{\alpha +\mu }%
\mathbf{Y}^{1}-\theta _{2}\frac{\sqrt{\alpha }F_{2}}{\alpha -\mu }\mathbf{Y}%
^{1}\right) \left[ \left( \mathbf{\hat{G}}^{\left( 0\right) }\right) ^{-1}%
\right] _{s=\mu }-  \label{75} \\
&&-\theta _{1}\theta _{2}\frac{\alpha }{\alpha ^{2}-\mu ^{2}}F_{1}F_{2}\left[
\left( \mathbf{\hat{G}}^{\left( 0\right) }\right) ^{-1}\right] _{s=\mu }%
\text{ },  \notag
\end{eqnarray}%
where%
\begin{equation}
\left[ \left( \mathbf{\hat{G}}^{\left( 0\right) }\right) ^{-1}\right]
_{s=\mu }=%
\begin{pmatrix}
\left( 2w\mu \right) ^{-p_{2}} & 0 \\ 
0 & \left( 2w\mu \right) ^{-p_{3}}%
\end{pmatrix}%
.  \label{76}
\end{equation}%
Now from formula (\ref{56}) we obtain the supervector $m^{Q}$: 
\begin{equation}
m^{Q}\left( \zeta ,\eta ,\theta _{1},\theta _{2}\right) =\left[ \mathbf{\hat{%
\Psi}}_{0}^{-1}\left( \zeta ,\eta ,\theta _{1},\theta _{2},s=\mu \right) %
\right] ^{QL}C_{L}\text{ }  \label{76-1}
\end{equation}%
with two arbitrary (in general complex) constants $C_{L}$.

7. For one soliton the supermatrix $\mathbf{X}$ from (\ref{58}) has only one
component 
\begin{equation}
(\mathbf{X)}_{11}=\frac{m^{Q}\left( \mathbf{\Psi }_{0}\right) _{QL}m^{L}}{%
\mu ^{2}-\alpha ^{2}}.  \label{76-2}
\end{equation}

8. Consequently the inverse supermatrix $\mathbf{X}^{-1}$ also has only one
component:%
\begin{equation}
\left( \mathbf{X}^{-1}\right) ^{11}=\frac{\mu ^{2}-\alpha ^{2}}{m^{Q}\left( 
\mathbf{\Psi }_{0}\right) _{QL}m^{L}}.  \label{77}
\end{equation}

9. Now from (\ref{65})-(\ref{66}) follows the dressed supermatrix $\mathbf{%
\Psi }(\zeta ,\eta ,\theta _{1},\theta _{2})$ [which is given by the
equation (\ref{25})] in terms of the all preceding findings:%
\begin{equation}
\mathbf{\Psi =\Psi }_{0}-\frac{\mu ^{2}-\alpha ^{2}}{\mu ^{2}T}\mathbf{\Psi }%
_{0}\left[ \left( \mathbf{\hat{\Psi}}_{0}\right) ^{-1}\right] _{s=\mu }%
\mathbf{C}\left[ \left( \mathbf{\hat{\Psi}}_{0}\right) ^{-1}\right] _{s=\mu }%
\mathbf{\Psi }_{0}\text{ },  \label{78}
\end{equation}%
where 
\begin{equation}
T=\text{Tr}\left\{ \left[ \left( \mathbf{\hat{\Psi}}_{0}\right) ^{-1}\right]
_{s=\mu }\mathbf{C}\left[ \left( \mathbf{\hat{\Psi}}_{0}\right) ^{-1}\right]
_{s=\mu }\mathbf{\Psi }_{0}\right\}  \label{79}
\end{equation}%
and matrix $\mathbf{C}$ has components $\left( \mathbf{C}\right) _{QL}$
formed by the products of the constants $C_{L}$:%
\begin{equation}
\left( \mathbf{C}\right) _{QL}=C_{Q}C_{L}\text{ }.  \label{80}
\end{equation}%
Calculating the coefficients in $\theta $-expansion of the right hand side
of the equation (\ref{78}) and equating this expansion to the expansion of
the left hand side $\mathbf{\Psi }$ we obtain the solutions for matrices $%
\mathbf{G},$ $\mathbf{G\Omega }_{1},$ $\mathbf{G\Omega }_{2},$ $\mathbf{GH}$%
. The first three of them follows from the free and linear terms of these $%
\theta $-expansions and they give the following final solutions for the
matrices $\mathbf{G},$ $\mathbf{\Omega }_{1},$ $\mathbf{\Omega }_{2}$ (only
which are of the interest):%
\begin{equation}
\mathbf{G=G}^{\left( 0\right) }-\frac{\mu ^{2}-\alpha ^{2}}{\mu ^{2}}\mathbf{%
G}^{\left( 0\right) }\mathbf{\Pi }\text{ },  \label{81}
\end{equation}%
\begin{gather}
\mathbf{\Omega }_{1}=\left( \mathbf{Y}^{1}+\frac{\mu ^{2}-\alpha ^{2}}{%
\alpha ^{2}}\mathbf{\Pi Y}^{1}\right) \frac{F_{1}}{\sqrt{\alpha }}+
\label{82} \\
+\frac{\alpha -\mu }{\mu \alpha ^{2}\left( k_{2}^{2}+k_{3}^{2}\right) }\left[
\mu \left( \alpha -\mu \right) \left( k_{2}^{2}-k_{3}^{2}\right) \mathbf{\Pi
+}2\alpha ^{2}\mathbf{\Gamma +}2\left( \mu ^{2}-\alpha ^{2}\right) \mathbf{%
\Pi \Gamma }\right] \frac{F_{1}}{\sqrt{\alpha }}\text{ },  \notag
\end{gather}%
\begin{gather}
\mathbf{\Omega }_{2}=\left( \mathbf{Y}^{1}+\frac{\mu ^{2}-\alpha ^{2}}{%
\alpha ^{2}}\mathbf{\Pi Y}^{1}\right) \frac{F_{2}}{\sqrt{\alpha }}+
\label{83} \\
+\frac{\alpha +\mu }{\mu \alpha ^{2}\left( k_{2}^{2}+k_{3}^{2}\right) }\left[
\mu \left( \alpha +\mu \right) \left( k_{2}^{2}-k_{3}^{2}\right) \mathbf{\Pi
-}2\alpha ^{2}\mathbf{\Gamma -}2\left( \mu ^{2}-\alpha ^{2}\right) \mathbf{%
\Pi \Gamma }\right] \frac{F_{2}}{\sqrt{\alpha }}\text{ }.  \notag
\end{gather}%
Here we used notations:

\begin{equation}
\mathbf{\Pi =}\frac{1}{k_{2}^{2}+k_{3}^{2}}%
\begin{pmatrix}
k_{2}^{2} & \alpha ^{1-2p_{2}}k_{2}k_{3} \\ 
\alpha ^{1-2p_{3}}k_{2}k_{3} & k_{3}^{2}%
\end{pmatrix}%
,\text{ \ }\mathbf{\Gamma =}%
\begin{pmatrix}
k_{2}^{2} & 0 \\ 
0 & -k_{3}^{2}%
\end{pmatrix}%
,  \label{84}
\end{equation}%
and%
\begin{equation}
k_{2}=\left( \frac{2w\mu }{\alpha }\right) ^{-p_{2}}C_{2}\text{ },\text{ \ }%
k_{3}=\left( \frac{2w\mu }{\alpha }\right) ^{-p_{3}}C_{3}\text{ }.
\label{85}
\end{equation}%
The direct calculation shows that the traces of the matrices $\mathbf{\Omega 
}_{1}$ and $\mathbf{\Omega }_{2}$ vanish in accordance with the requirements
(\ref{44}) and all four matrices $\mathbf{G},$ $\mathbf{G\Omega }_{1},$ $%
\mathbf{G\Omega }_{2},$ $\mathbf{GH}$ are symmetric as it should be.

10. To calculate the components of spinors $\chi ^{\left( 1\right) }(\zeta
,\eta )$ and $\chi ^{\left( 2\right) }(\zeta ,\eta )$ we need to fix some
frame $\mathbf{V.}$ Let's chose the triangular frame with zero at its lower
left place. The one-soliton matrix $\mathbf{G}$ (\ref{81}) with notations (%
\ref{84}) and (\ref{85}) take the form:%
\begin{equation}
\mathbf{G=}%
\begin{pmatrix}
\frac{\alpha ^{2}k_{2}^{2}+\mu ^{2}k_{3}^{2}}{\mu ^{2}\left(
k_{2}^{2}+k_{3}^{2}\right) }\alpha ^{2p_{2}} & \frac{\alpha \left( \alpha
^{2}-\mu ^{2}\right) k_{2}k_{3}}{\mu ^{2}\left( k_{2}^{2}+k_{3}^{2}\right) }
\\ 
\frac{\alpha \left( \alpha ^{2}-\mu ^{2}\right) k_{2}k_{3}}{\mu ^{2}\left(
k_{2}^{2}+k_{3}^{2}\right) } & \frac{\alpha ^{2}k_{3}^{2}+\mu ^{2}k_{2}^{2}}{%
\mu ^{2}\left( k_{2}^{2}+k_{3}^{2}\right) }\alpha ^{2p_{3}}%
\end{pmatrix}%
\text{ },\text{ \ }\det \mathbf{G=}\frac{\alpha ^{4}}{\mu ^{2}}\text{ }.
\label{86}
\end{equation}%
The triangular frame, satisfying relation $\mathbf{G}=\mathbf{V\tilde{V},}$
is:%
\begin{equation}
\mathbf{V=}\left[ \frac{\alpha ^{2}k_{3}^{2}+\mu ^{2}k_{2}^{2}}{\mu
^{2}\left( k_{2}^{2}+k_{3}^{2}\right) }\alpha ^{2p_{3}}\right] ^{-1/2}%
\begin{pmatrix}
\frac{\alpha ^{2}}{\mu } & \frac{\alpha \left( \alpha ^{2}-\mu ^{2}\right)
k_{2}k_{3}}{\mu ^{2}\left( k_{2}^{2}+k_{3}^{2}\right) } \\ 
0 & \frac{\alpha ^{2}k_{3}^{2}+\mu ^{2}k_{2}^{2}}{\mu ^{2}\left(
k_{2}^{2}+k_{3}^{2}\right) }\alpha ^{2p_{3}}%
\end{pmatrix}%
\text{ }.  \label{87}
\end{equation}%
Forming from this $\mathbf{\tilde{V}}$ and\textbf{\ }$\mathbf{\tilde{V}}%
^{-1} $ and using them to calculate matrices $\mathbf{\tilde{V}\Omega }_{1}%
\mathbf{\tilde{V}}^{-1}$ and $\mathbf{\tilde{V}\Omega }_{2}\mathbf{\tilde{V}}%
^{-1}$ [with $\mathbf{\Omega }_{1}$ and $\mathbf{\Omega }_{2}$\textbf{\ }%
from (\ref{82})-(\ref{83})] we get from (\ref{43-1}) the final result for
the components of the spinors (\ref{40}). They can be represented in the
form:%
\begin{equation}
a_{1}+b_{1}=\varkappa \left[ 1-\frac{2\left( \alpha -\mu \right)
^{2}k_{2}^{2}k_{3}^{2}}{\left( \mu ^{2}k_{2}^{2}+\alpha ^{2}k_{3}^{2}\right)
\left( k_{2}^{2}+k_{3}^{2}\right) }\right] \frac{F_{1}}{\sqrt{\alpha }},
\label{88}
\end{equation}%
\begin{equation}
a_{1}-b_{1}=\varkappa \left[ 1-\frac{2\left( \alpha +\mu \right)
^{2}k_{2}^{2}k_{3}^{2}}{\left( \mu ^{2}k_{2}^{2}+\alpha ^{2}k_{3}^{2}\right)
\left( k_{2}^{2}+k_{3}^{2}\right) }\right] \frac{F_{2}}{\sqrt{\alpha }},
\label{89}
\end{equation}%
\begin{equation}
a_{2}+b_{2}=\varkappa \left[ \frac{2\left( \alpha -\mu \right)
k_{2}k_{3}\left( \mu k_{2}^{2}+\alpha k_{3}^{2}\right) }{\left( \mu
^{2}k_{2}^{2}+\alpha ^{2}k_{3}^{2}\right) \left( k_{2}^{2}+k_{3}^{2}\right) }%
\right] \frac{F_{1}}{\sqrt{\alpha }},  \label{90}
\end{equation}%
\begin{equation}
a_{2}-b_{2}=\varkappa \left[ \frac{2\left( \alpha +\mu \right)
k_{2}k_{3}\left( \mu k_{2}^{2}-\alpha k_{3}^{2}\right) }{\left( \mu
^{2}k_{2}^{2}+\alpha ^{2}k_{3}^{2}\right) \left( k_{2}^{2}+k_{3}^{2}\right) }%
\right] \frac{F_{2}}{\sqrt{\alpha }},  \label{91}
\end{equation}%
where 
\begin{equation}
\varkappa =\left( 3i\right) ^{-1/2}2^{3/4}.  \label{92}
\end{equation}

11. Finally we should replace $\mathbf{G}$ (\ref{86}) by its "physical" value%
\begin{equation}
\mathbf{G}_{ph}=\frac{\mu }{\alpha }\mathbf{G}  \label{93}
\end{equation}%
which also is the solution of the equation (\ref{33}) and plus to this
satisfy the condition $\det \mathbf{G}_{ph}=\alpha ^{2}$ (correspondingly we
should pass to the "physical" frame $\mathbf{V}_{ph}=\left( \mu /\alpha
\right) ^{1/2}\mathbf{V}$). Then we have to use namely $\mathbf{G}_{ph}$ as
final solution for the $\mathbf{G}$-matrix. The solutions for the matrices $%
\mathbf{\Omega }_{1}$ and $\mathbf{\Omega }_{2}$ and for the spinors remains
unchanged due to the invariance of the equations (\ref{34}), (\ref{35}) and (%
\ref{43-1}) under this operation.

The necessary reality conditions can be provided by the appropriate choice
of the arbitrary constants $C_{2},C_{3},w$ and suitable choice of the
solution (\ref{74}) for the pole trajectory $\mu $.

Finally it is worth mentioning that solution constructed here although being
oversimplified and serving only for the demonstration of the ability of
apparatus represents nevertheless some physical interest. Such kind of
solutions correspond to that approximation when one are going to find
behavior of spinors in the fixed external field neglecting by their back
reaction. The number of solitons in the external field $\mathbf{G}$ can be
arbitrary. For example, we can arrange two-solitonic solution for this field
and in this case (after complex transformation to the stationary $\mathbf{G}$
with axial symmetry \cite{BZ1}) we can describe the behavior of spinors in
the external field of the Kerr and Schwarzschild black holes. It is
interesting that for the such particular cases one can abstract oneself from
the supersymmetry and anticommuting properties of the spinorial variables.
However, the technique described in this paper permits to build the exact
and complete solutions taking into account also the full non-linear
fermionic terms in equations (\ref{9})-(\ref{10}) and in such cases spinors
produce "souls" in the bosonic fields which is sophisticated phenomenon
having no clear interpretation in the realm of classical physics.

\textbf{ACKNOWLEDGEMENT}. It is my pleasure to express the gratitude to
Hermann Nicolai for illuminating discussions, critics and valuable comments
as well as for his hospitality at Albert Einstein Institute at Golm, where
part of this work has been done.


\begin{thebibliography}{9}
\bibitem{M} D. Maison "Are the Stationary, Axially Symmetric Einstein \
Equations Completely Integrable?", Phys. Rev. Lett. \textbf{41}, 521 (1978).

\bibitem{BZ1} V.A. Belinski and V.E. Zakharov "Integration of the Einstein
equations by means of the inverse scattering problem technique and
construction of exact soliton solutions", Sov. Phys. JETP \textbf{48}, 985
(1978).

\bibitem{N1} H. Nicolai "The integrability of N=16 supergravity", Phys.
Lett. \textbf{B194}, 402 (1987).

\bibitem{N2} H. Nicolai "Two-Dimensional Gravities and Supergravities as
Intgrable Systems", Lectures Notes in Physics, vol. \textbf{396}, p. 231
(1991).

\bibitem{Mik} A.V. Mikha\u{\i}lov "Integrability of supersymmetrical
generalization of classical chiral models in two-dimensional space-time",
JETP Lett., \textbf{28}, 512 (1978).
\end{thebibliography}
\end{document}